# An Overview of the KamLAND 1-kiloton Liquid Scintillator *

F. Suekane**, T. Iwamoto, H. Ogawa,
O. Tajima and H. Watanabe,
(for the KamLAND RCNS Group***)
***Research Center for Neutrino Science (RCνS),***
***Tohoku University, Sendai, 980-8578, Japan***

## Abstract

KamLAND (Kamioka Liquid Scintillator Antineutrino Detector) is a massive liquid-scintillator-based neutrino detector studying low energy neutrino oscillation. The experiment has discovered a large deficit of the reactor neutrino flux with baseline of 180km in the year 2002. All the solar neutrino solutions, except for the Large Mixing Angle solution, are excluded with high confidence level. Assuming the deficit is caused by the oscillation phenomenon of the electron type neutrinos, the most likely oscillation parameters are measured to be $\sin^2 2\theta = 1$ and $\Delta m^2 = 6.9\times10^{-5} eV^2$. These results are placing a strong impact on the elementary particle physics. The KamLAND detector makes use of one kiloton of liquid scintillator (LS) as the active neutrino target. This is the largest homogeneous LS detector in the world. Although the light yield and attenuation length of the LS were measured to be 57% anthracene and ~10m, in the test experiments, when it is used in the KamLAND detector, the effective transparency and light yield significantly increased from these values, presumably due to the scattering and re-emission in the large scale LS. The radio active contamination in the LS is extremely low, such as $3.5\times10^{-18} g/g$ for uranium, $5.2\times10^{-17} g/g$ for thorium and $< 2.7\times10^{-16} g/g$ for $^{40}$K. This is the world's lowest contamination record so far achieved and measured. The success of the KamLAND experiment is a fruit of this high-quality and large-quantity LS. In this proceedings, an outline of the LS is described. A technical paper which describes the LS in detail is now being prepared.

---

** The speaker. Electric address: suekane@awa.tohoku.ac.jp

*** KamLAND RCNS members since the R&D of LS started (some of them have moved.): T. Araki, K. Eguchi, S. Enomoto, K. Furuno, H. Hanada, A. Hasegawa, S. Hatakeyama, G. Horton-Smith, K. Ichiumura, H. Ikeda, K. Ikeda, K. Inoue, K. Ishihara, T. Itoh, W. Itoh, T. Iwamoto, T. Kawaguchi, S. Kawakami, T. Kawashima, T. Kinebuchi, H. Kinoshita, M. Koga, Y. Koseki, T. Maeda, K. Mashiko, T. Mitsui, M. Motoki, K. Nakajima, M. Nakajima, T. Nakjima, I. Nishiyama, H. Ogawa, K. Oki, K. Owada, M. Saitoh, T. Sakabe, I. Shimizu, J. Shirai, F. Suekane, A. Suzuki, K. Tada, K. Tagashira, O. Tajima, D. Takagi, T. Takayama, K. Tamae, Y. Tsuda and H. Watanabe.

## 1 Introduction

The properties of neutrinos are least known among the standard fermions due to the extreme difficulty of its detection. However, the experimental studies of neutrinos are progressing very rapidly in these days thanks to the progress of neutrino detection technologies. The KamLAND experiment has been playing major roll among such experiments since its discovery of the deficit of the reactor neutrinos[1]. KamLAND is the neutrino detector which is designed to detect low energy neutrinos by using 1kiloton of liquid scintillator. The main purpose of this experiment is to search for oscillation phenomena of neutrino by detecting reactor $\bar{\nu}_e$ which comes from hundreds kilometers away. The typical energy of the reactor $\bar{\nu}_e$ is a few MeV. There are a number of reactors at a mean distance of 180km from KamLAND and the expected event rate of $\bar{\nu}_e$ coming from such reactors is about 2 per day per kiloton if there is no neutrino oscillation. If neutrino oscillation exists, the event rate reduces to;

$$N_\nu(E_\nu) = N_\nu^0(E_\nu)\left(1-\sin^2 2\theta \sin^2 \frac{\Delta m^2 L}{4E_\nu}\right)$$

where, $E_\nu$ is the neutrino energy. $\theta$ is the mixing angle between the mass eigenstates and the weak eigenstates. $\Delta m^2$ is the difference of mass squared of two type neutrinos. $N_\nu(E_\nu)$ and $N_\nu^0(E_\nu)$ are the neutrino event rate with and without neutrino oscillation, respectively. $L$ is the distance between neutrino source and the detector. It is very important to study neutrino oscillation because, if neutrino oscillation exists, it means that at least one neutrino has finite mass and that neutrino states mix. It is possible to access to very low mass scale through neutrino oscillation for which any other practical direct measurements can not access to. The values of the neutrino mass relations and mixing angles will play an important role when constructing the unified theory of elementary particles.

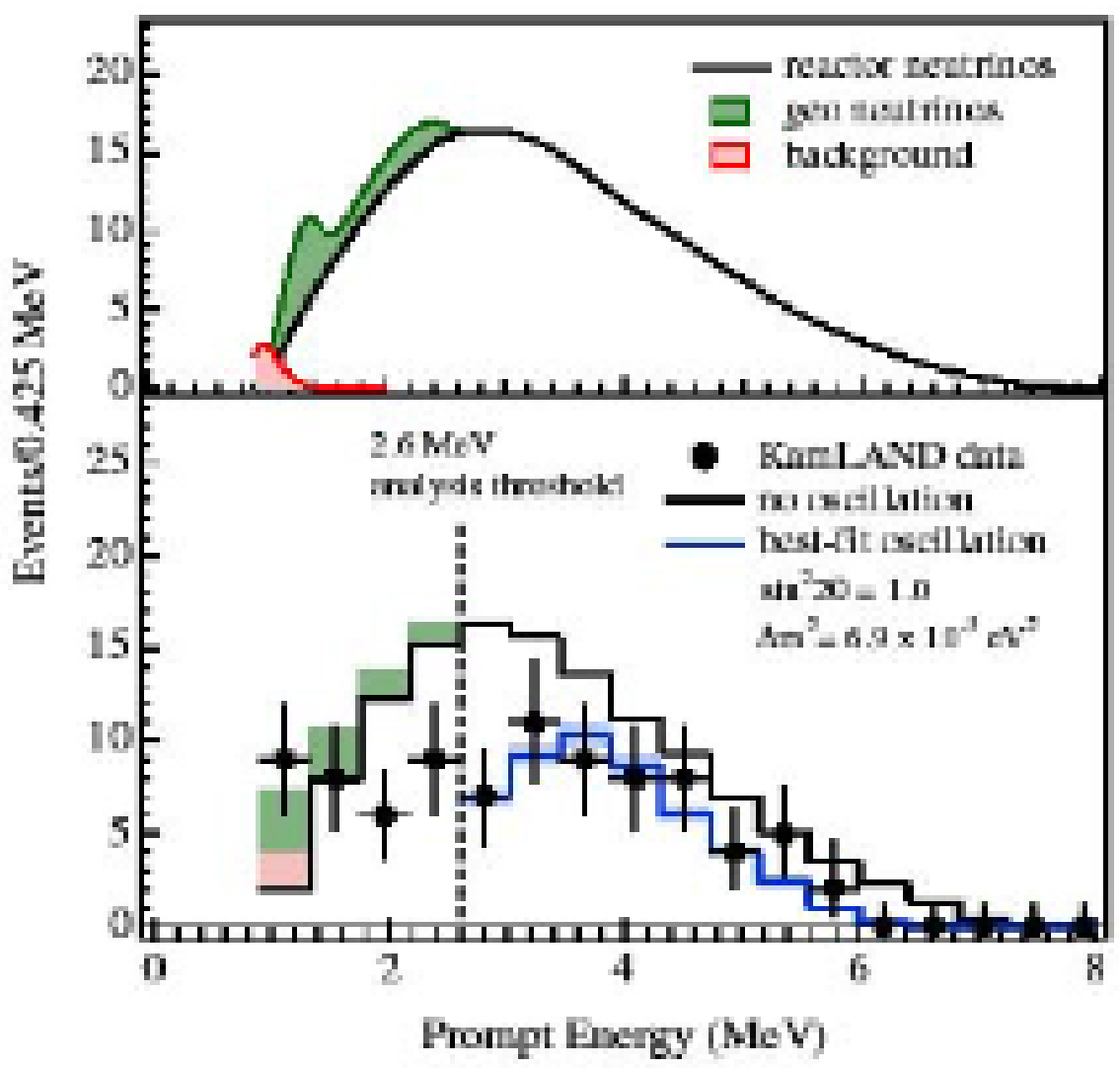

Fig.-1: The neutrino energy spectrum. The horizontal axis corresponds to neutrino energy minus 0.8MeV.

The KamLAND project was approved in 1997 and the detector construction was finished in the year 2001 and started data taking in January 2002. Within the year 2002, the KamLAND group found 39% of large deficit in the reactor neutrino flux and settled down the long standing solar neutrino problem [2] with man-made neutrinos for the first time. Fig.-1 shows the energy spectrum of the reactor $\bar{\nu}_e$ events. A large deficit in the event rate is clearly seen. Assuming that the deficit is caused by neutrino oscillation, the most likely oscillation parameters are measured to be $\sin^2 2\theta = 1$ and $\Delta m^2 = 6.9\times10^{-5} eV^2$ . These results were published in December 2002 [1] and have been placing strong impacts on elementary particle physics. Recently KamLAND searched for $\bar{\nu}_e$ from the sun and set the strongest upper limits on the flux [3]. In near future, KamLAND is expected to publish the analysis result of geo-$\bar{\nu}_e$ [4] detection. The success of such measurements is indebted to the success of the development of the high quality and large quantity liquid scintillator for the KamLAND detector.

## 2. The KamLAND Detector

Fig.-2 shows the schematic view of the KamLAND detector. It consists of, from inner to outer, 1,200m$^3$ liquid scintillator (LS) as active neutrino target, 13m-diameter balloon made of transparent Nylon/EVOH(Ethylene Vinyl Alcohol) film to hold the LS, 1,800m$^3$ of buffer oil to cancel the LS weight and to shield γ-ray background, 1,325 17 inch-aperture PMT and 554 20 inch PMT to measure the scintillation light, 18m-diameter stainless steel tank, and cosmic-ray anti-counter made of 3,000m$^3$ pure water. The detector locates 1,000m below the top of the Ikenoyama mountain to reduce the cosmic rays to $10^{-5}$-fold. The PMT photo coverage is 22% for 17 inch PMTs only and 34% with 20 inch PMTs. The quantum efficiency of the PMTs is about 20% for 340~400nm light.

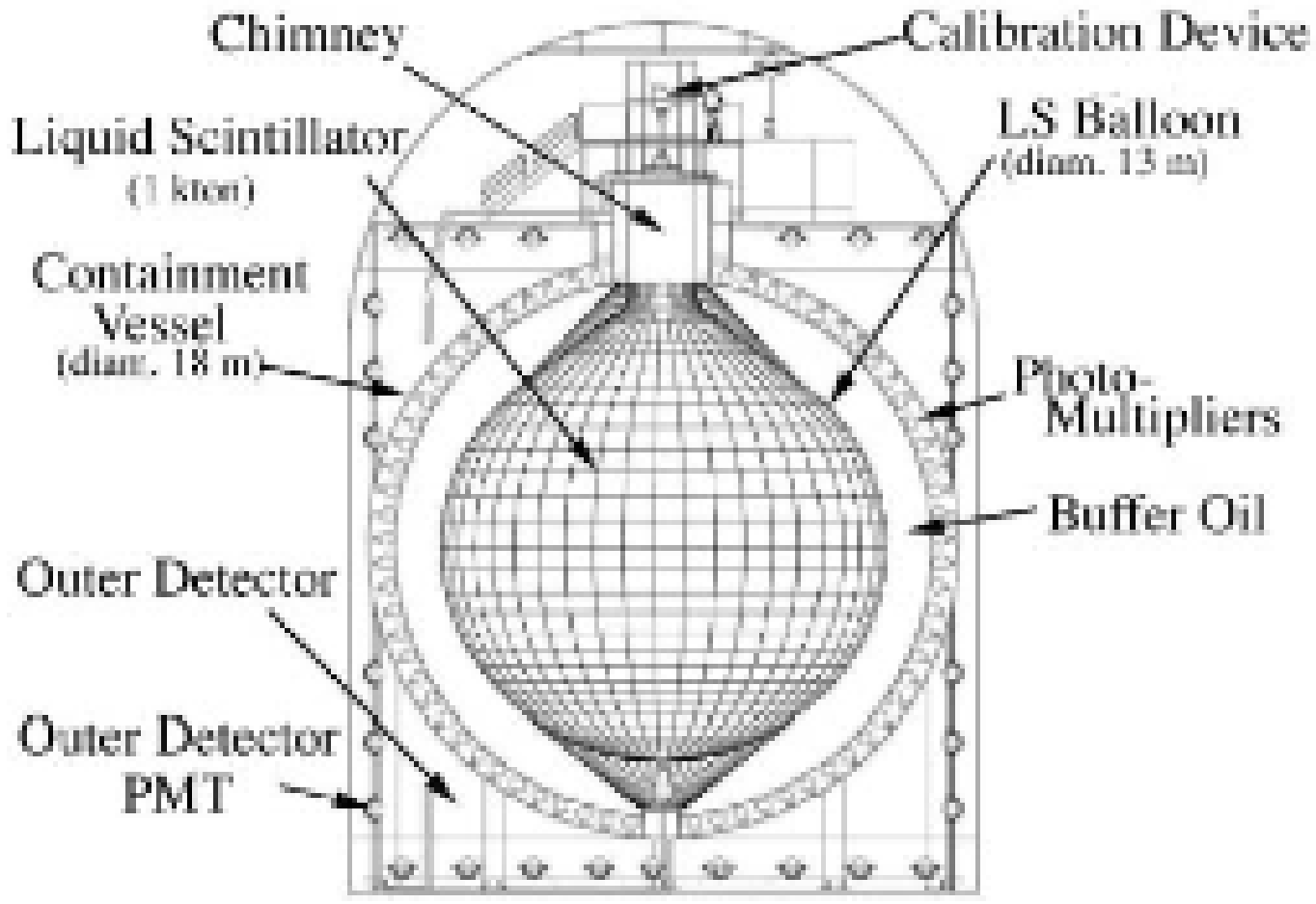

Fig.-2: The KamLAND Detector.

The KamLAND detector detects $\bar{\nu}_e$ coming from distant reactors whose typical energy is a few MeV. When reactor $\bar{\nu}_e$ performs charged current interaction, colliding with protons, it turns to positron and the proton turns to neutron.

$$\bar{\nu}_e + p \rightarrow e^+ + n$$

The positron emits scintillation lights when it goes through the LS. On the other hand, the neutron quickly thermalises by colliding with protons around and is eventually absorbed by a proton, emitting 2.2MeV γ-ray.

$$n + p \rightarrow d + \gamma\,(2.2MeV)$$

The neutron absorption occurs typically 200μs after the positron signal and the γ-ray produces mono-energetic delayed signal. By requiring the existence of the two correlated signals, the accidental backgrounds are severely suppressed. The expected event rate is 2/day for whole LS volume in case of no neutrino oscillation. Liquid scintillator was chosen as the active neutrino target because, it is possible to construct large and homogeneous detector, the radio active background level is intrinsically very low in petroleum based oils and it is very much cost effective. The light output of LS is typically several tens of times of that of Cherenkov light. Thus it is possible to detector low energy signals for which Cherenkov-base detector can not reach.

## 3. The Liquid Scintillator

The KamLAND LS was developed by KamLAND RCNS group and there have been a number of reports about it [5]. The KamLAND LS consists of

<u>80v% of normal-Dodecane + 20v% of Pseudocumene+1.52g/liter of PPO,</u>

where, v% stands for % in volume. Fig.-3 shows molecular structure of these chemicals.

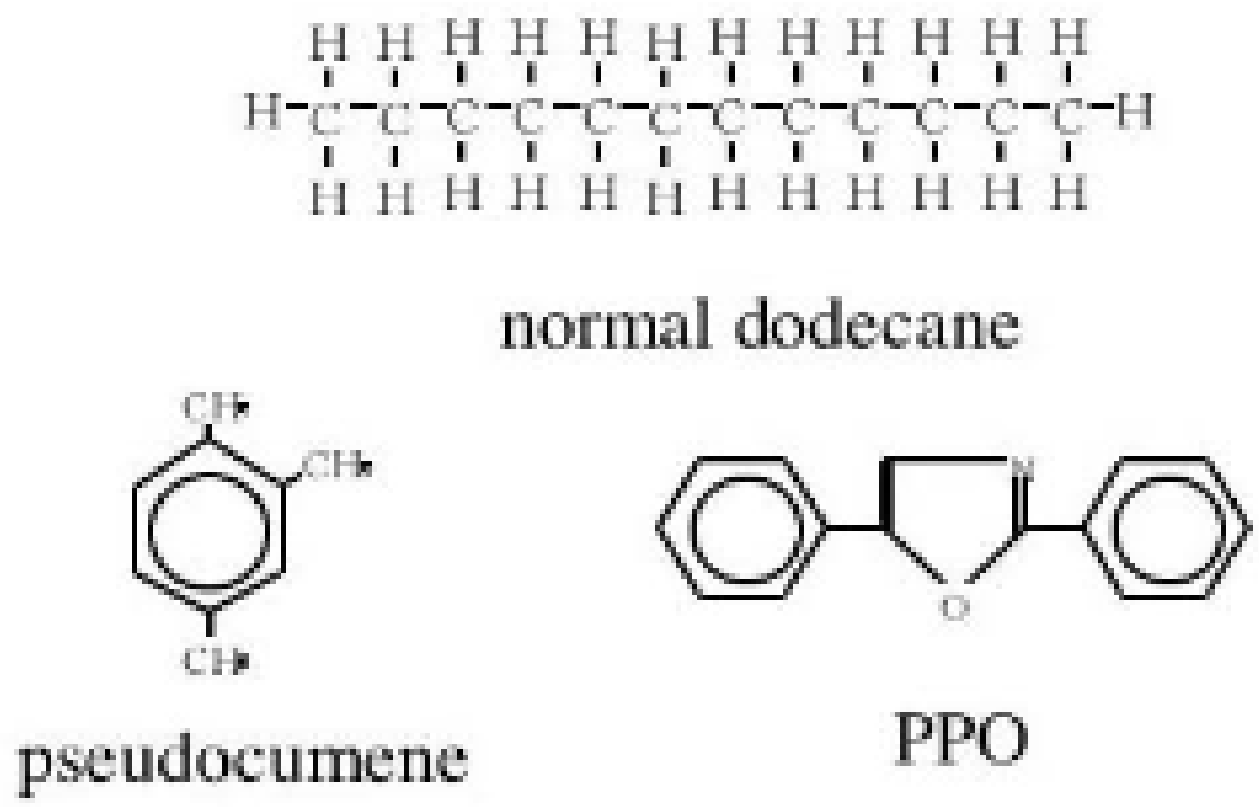

Fig.-3: Chemicals used in the KamLAND LS.

PPO(2,5-Diphenyloxazole) is a flour which has been widely used in large scale liquid scintillators for high energy physics and neutrino experiments[6]. Fig.-4 shows the emission spectra of PPO and quantum efficiency of KamLAND PMT cathode. The photo cathode of the PMT is a bi-alkali type, which has maximum sensitivity for 360nm wavelength light. This figure shows that the matching of PPO emission spectrum and high sensitive region of the PMT cathode is good and no secondary wave shifter is necessary. PPO is the most 'dirty' material used in the KamLAND LS. KamLAND uses PPO which was extra-purified with additional sets of crystalizations by the supply company. It is further purified on site by water extraction by KamLAND group.

Pseudocumene (PC) is an aromatic solvent whose chemical name is 1,2,4-trimethylbenzene. Aromatic solvents give LS high light output and neutron/***g*** and $\alpha/\gamma$ pulse shape discrimination capability. PC has a high flash point and low chemical reactivity among the BTXC (Benzene, Toluene, Xylene, Cumene) families. Pure PC is transparent to PPO scintillation light. The PC concentration was determined by optimizing the flash point, light output and transparency for the purpose of our experiment. PC is produced by fine distillation after the desulfurization and hydrogenation of the petroleum oil. PC is industrially produced and it is possible to obtain large amount of high quality PC in a short time period with reasonable price.

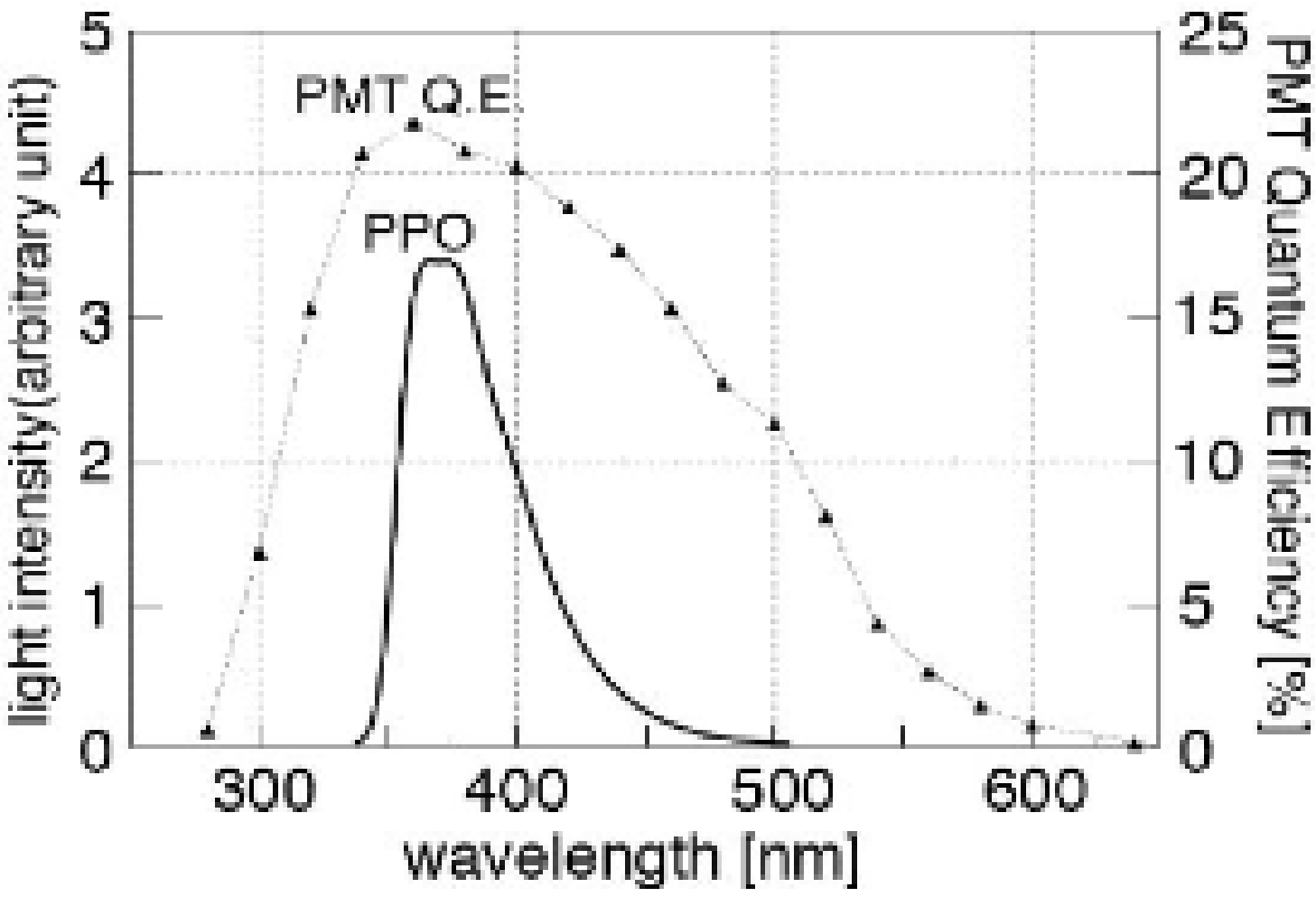

Fig.-4: Emission spectra of PPO and quantum efficiency of photo cathode.

Normal-dodecane (ND) does not have double chemical bonding nor circular structure. It means that they have no absorption at scintillation light and is chemically stable and immune to oxidization. ND has significantly less reactivity than PC and it is safer for balloon material. The PC+PPO liquid scintillator is diluted with ND to obtain better transparency, higher flash point and better stability. Because detector size is large, a very good transparency is essential to obtain high, stable and uniform light detection efficiency. ND has higher flash point (83°C) than PC (54°C) and adding ND, the safety of the LS improved significantly. ND is produced by refining from the petroleum oil. Firstly normal-paraffins are selected from oils by using molecular sieves after desulfurization and hydrogenation. Then ND is separated from other normal paraffins by the fine distillations. The temperature window for the distillation is very narrow and contamination is generally very low level. The ND is selected among other normal paraffins because it has high flash point, yet reasonably high melting point. ND is used in the nuclear industry and a large amount of high quality ND is available with reasonable price. The properties of KamLAND Liquid Scintillator are summarized in Table-1. In the table, 'Effective' means that the property includes the effect of scattering and re-emission of the scintillation light in the KamLAND detector. This effect could not be studied in the small test experiments.

## 4. The Liquid Scintillator R&D

Probably the KamLAND LS is one of the most simple liquid scintillators. It does not contain secondary wavelength shifter, antioxidant nor any hidden ingredients. It is the result of our strategy for the LS development. There is a special difficulty for the KamLAND detector, that is the LS can not be replaced once it is filled in the detector because there is no reservoir tanks. So that we had to be absolutely confident that the LS had enough properties before filling in the detector. Also the deadline had been already set when the R&D was started in 1997. Based on these conditions, the strategy of the LS R&D was to use only the proven technologies and to make the scintillator components as simple as possible. The aim of the test experiments was to verify the tested property satisfies the requirement or not, not to evaluate the quality in detail nor to make thorough investigations. The LS is developed under the following requirements. (1) The energy resolution is to be better than $10\%/\sqrt{E(MeV)}$ to perform the spectrum analysis of the neutrino energy. In order to achieve this resolution, the light output should be larger than 40% anthracene and the light attenuation length should be longer than 10m. (2) In order to achieve sufficient signal to noise ratio, the U/Th/K contents should be less than $10^{-14}/10^{-14}/10^{-12}$g/g, respectively. (3) For safety reason, the flash point should be higher than 60°$C$. (4) The LS should be stable for several years. (5) Large amount of high quality LS ingredients are to be supplied within half year period. And (6) The price should be less than a few dollars/liter. Many sorts of test measurements had been performed before the final LS components were determined. Some examples of the test items were, light outputs, emission spectra, absorption spectra, transparency with optical photometer, transparency with 3m long tube, radio purities, a quench, oxygen quench, n/γ and α/γ PSD, stability, aging, compatibility with materials, 6.3m path length direct light yield measurement, 1m$^3$ test bench experiment, flash point, possibility of bacteria breeding, efficiency of purification, efficiency of $N_2$ bubbling, etc. It is impossible to describe all of these items in the limited page of this proceedings and only the direct light yield measurements is described below. The detail of the LS R&D will be reported elsewhere.

Table-1: KamLAND Liquid Scintillator properties

| Item | Property |
|---|---|
| Direct Light Output | 57% anthracene |
| Effective light output | ~70% anthracene* |
| Dirext Light Attenutaion Length | 10m (@λ=400nm) |
| Effective light attenuation length | ~20m* |
| Energy Resolution in KamLAND | $7.5\%/\sqrt{E(MeV)}$ |
| Specific Gravity | 0.778g/cm$^3$ |
| Flash Point | 64°$C$ |
| Pulse Decay Time | 6ns |
| Refractive Index | 1.44 (@λ=589nm, 15°$C$) |
| H/C ratio | 1.97 |
| Radio Purity (radio equilibrim assumed) | U: $3.5\times10^{-18}g/g$ |
| | Th: $5.2\times10^{-17}g/g$ |
| | $^{40}$K: $<3.7\times10^{-16}g/g$ |
| α quench factor | 13 for 7.7MeV α |

* Preliminary Value

One possible problem of the R&D is that, in general, the size of the test experiments are very much smaller than the actual KamLAND detector and there is ambiguity when extrapolating the performance of the test experiments to the real detector size. So we did the final checkout by one-dimensional real size test experiment to make sure that the photoelectron yield is enough after passing through 6.4m of LS. Figs.-5 show the measurement apparatus and the measured pulse shape. 100-liter of LS was filled in the 6.83m long black anodized Aluminum pipe whose inner diameter is 14cm. Both ends of the pipe are sealed by 35mm thick UV transparent acrylic plates and viewed by 5" PMTs with same photo cathode material as the KamLAND PMT. The scintillation light was generated by the cosmic-ray muons which pass across one end of the scintillation pipe. Three plastic scintillators sandwiched the scintillator pipe and triggered the cosmic rays. The lead brick between the trigger counters was to select high energy cosmic rays. The PMT at the near end with respect to the trigger counters measured the energy deposit of the cosmic rays.

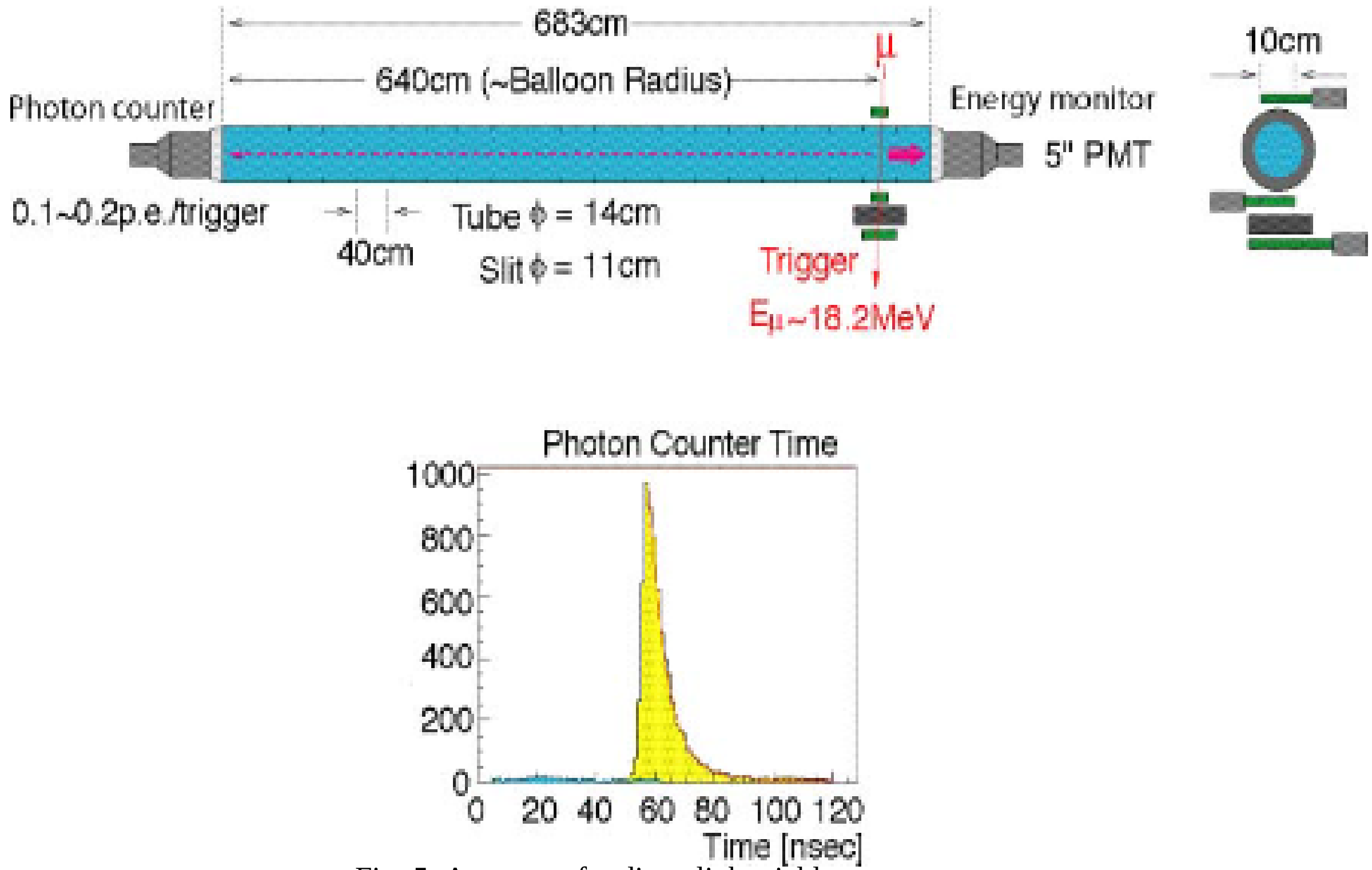

Fig.-5: Apparatus for direct light yield measurement and measured timing spectrum.

The PMT at the far end counts the number of photoelectrons and measures the time since the trigger signal. In order to suppress the light reflection from the inner wall of the pipe, there were thin black anodized aluminum rings at intervals of 40cm. The inner diameter of the rings was 11cm and the thickness was 2mm. More than 99.99% of reflections in the pipe was suppressed by this structure and only the direct lights were detected in the photon counter PMT. The distance between the trigger counters and the photon counter PMT was 640cm, which corresponds to the radius of the scintillator balloon of 650cm. The solid angle of the photon counter PMT viewed from the cosmic rays was $2.3\times 10^{-4}$ sr. The mean energy deposit by cosmic ray was around 15MeV and the expected number of photoelectrons in the photon counter PMT was 0.1~0.2p.e. /trigger. Therefore the direct light yield can be obtained by the photon counting method. The timing structure of the scintillation emission also can directly be measured from the TDC distribution. The energy calibration was performed by using the Compton-scattered electron caused by the back scattered 662KeV $^{137}Cs$ γ-ray tagged by a NaI counter behind the source. The lower figure of the Figs.-5 shows the timing distribution of the photon counter PMT. A beautiful pulse shape was observed and the decay time of the scintillation was obtained from the tail of the shape to be around 6ns. The light yield of the LS was calculated from the ratio of the number of events in the photon counter PMT and the number of cosmic-ray triggers. The light yields were measured by changing the PPO concentration. Fig.-6 shows the result of the measurement. For low PPO concentrations, the light output of the LS is small and the photoelectron yield is also small. On the other hand, at too high PPO concentrations, the self absorption and scattering of the scintillation light start to dominate and the total photoelectron yield reduces.

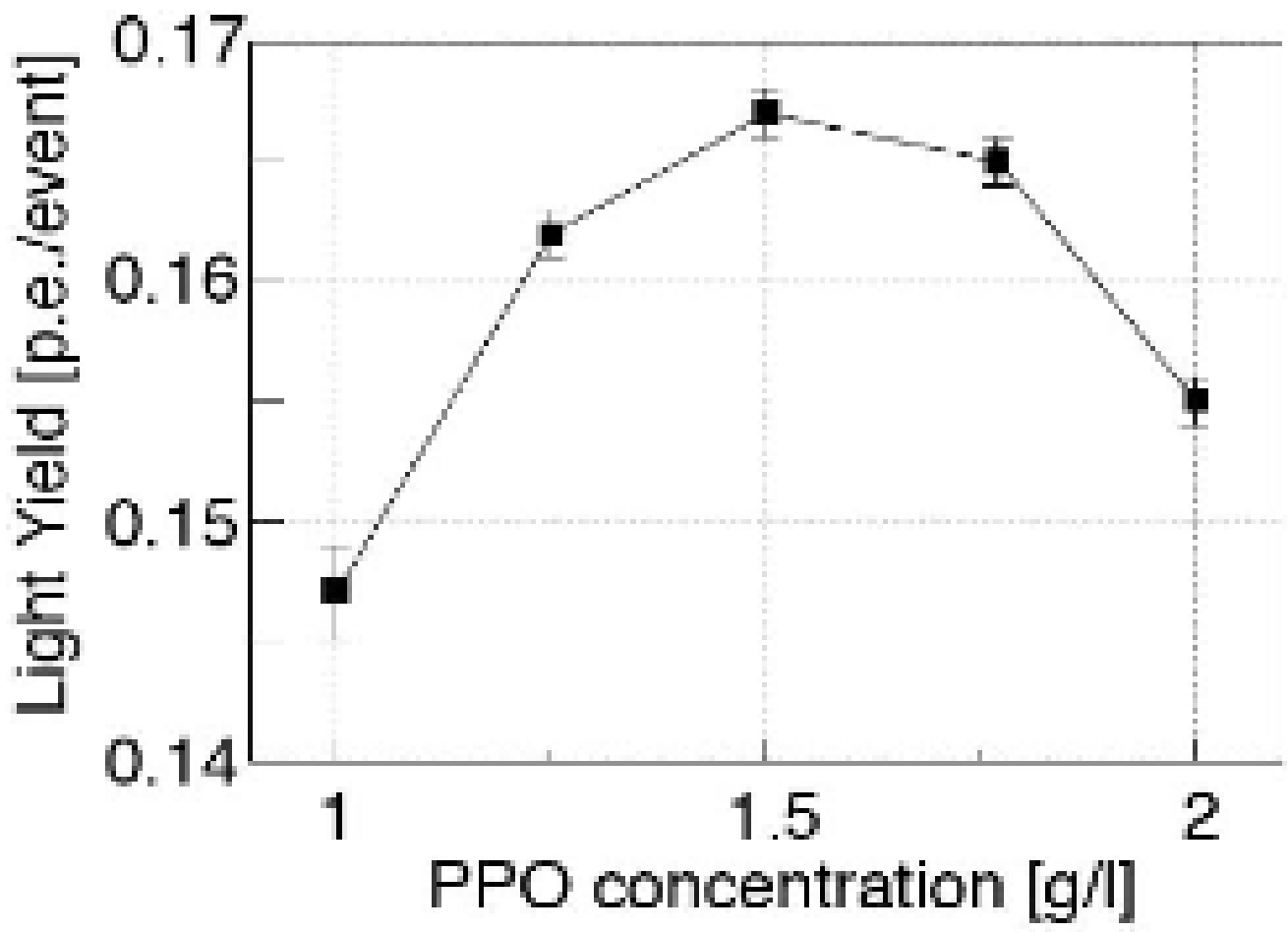

Fig.-6: The Light yield after passing through 6.4m for various PPO concentrations.

The photoelectron yield became maximum at 1.5g/liter of PPO, where the expected photoelectron yield in the KamLAND detector was calculated to be 200p.e./MeV. This satisfies the requirement of >100p.e./MeV and we became confident that this LS could be used in the actual KamLAND detector.

## 5. Formulation of the Liquid Scintillator

The LS formulation took place from May 2001 to September 2001. The blending of LS were performed in the oil purification system at the site of the KamLAND experiment in the mine. 1,800m$^3$ of ND, 1,000m$^3$ of isoparaffin (IP), 240m$^3$ PC and 1.8tons of PPO are used in the KamLAND detector. The IP and a part of ND are used for the buffer oil (BO). PPO had been delivered from Europe and dissolved in PC with high concentration and purified with water extraction beforehand. Five stainless steel lorry tracks with 12m$^3$ loading capacity carried ND and IP into the mine every day and the oils were unloaded at the KamLAND purification system and stored in stainless steel tanks. ND was delivered from a petrochemical complex in Mizushima and IP, from Yokohama. The PC was delivered from Matsuyama using stainless steel drum cans. About 230 lorry deliveries and net 1,200 drum cans were needed. The pure PC and the PPO-concentrate PC were introduced into the purification system using a handy drum pump. Then the liquids in the purification system was self-circulated to blend and formulate the LS and the BO. The formed LS and BO were purified and sent to the balloon and the stainless steel tank every day after the purification. The filling finished in September 2001 and data taking started from January 2002 after the installation of electronics.

## 6. The Performances of LS in the KamLAND Detector

Fig.-7 shows an example of event display. The observed photo-electron yield in the KamLAND detector is approximately 300p.e./MeV. This value is much higher than the expectation based on the result of the direct light yield measurement. This increase can be explained by the scattering and the re-emission of the scintillation light in the large scintillator volume. The transparency is being studied by the dependence of the photo-electron yield on the distance between PMTs and calibration

sources deployed at known positions. The preliminary value of the effective transparency is roughly 20m, where the 'effective' means that the scattering and re-emission light in the LS are contained in the measurement. The waveform of each PMT is recorded and the information is converted to the timing and charge in the data analysis. The energy is calculated based on the number of photoelectron and the position is reconstructed based on the timing.

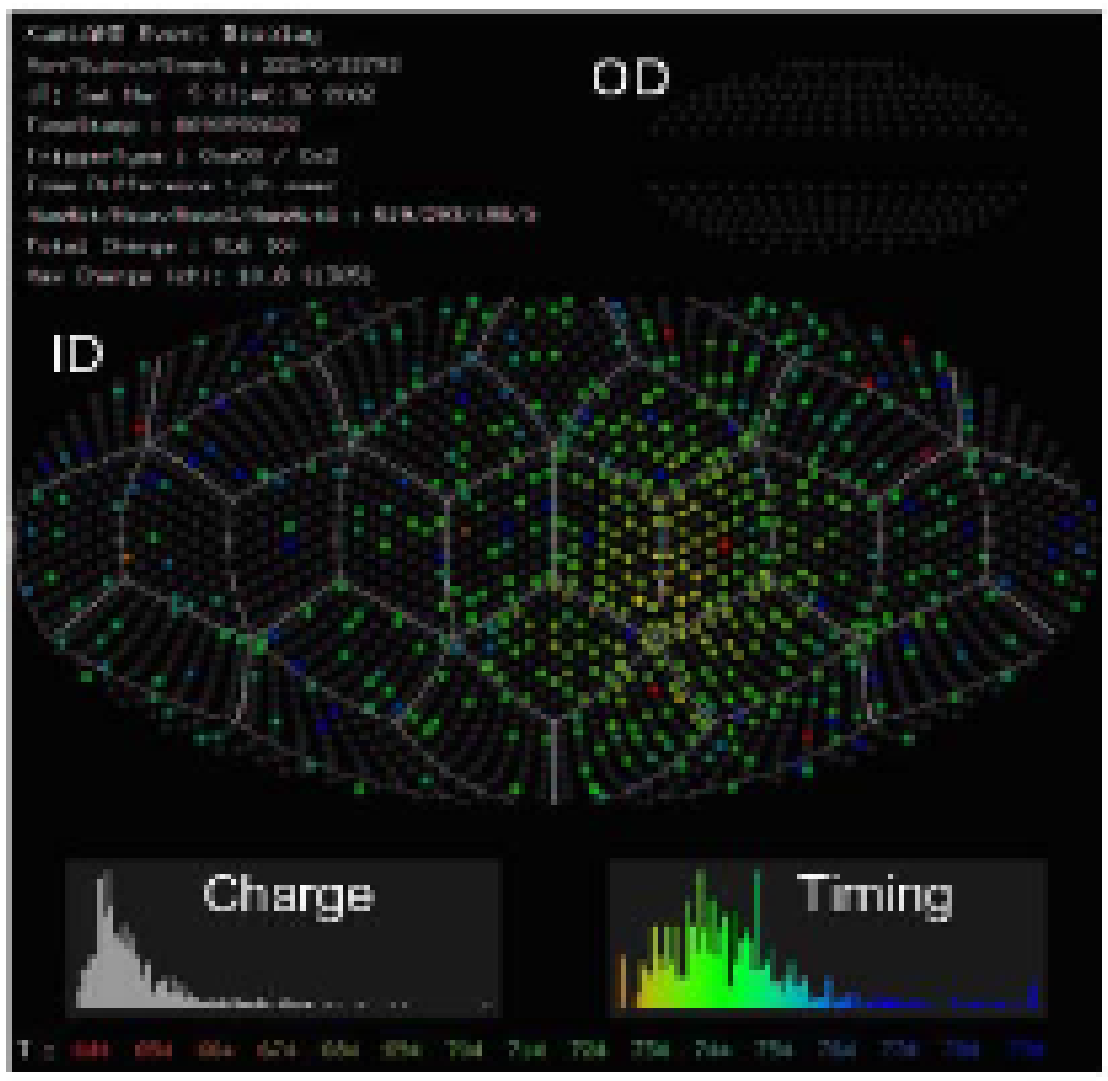

Fig.-7: Event display

The energy and position calibrations have been done by introducing $^{60}Co$(2.5MeV), $^{65}Zn$(1.116MeV) and $^{68}Ge$(1.022MeV) in the LS from the chimney of the stainless steel tank. Neutron absorption signals on proton (2.2MeV) and $^{12}C$ (4.95MeV) add additional calibration points. Fig.-8 shows the energy distributions for such signals. The energy resolution is measured to be $7.5\%/\sqrt{\boldsymbol{E}(\boldsymbol{MeV})}$ , which is much better than the requirement due to higher photoelectron statistics.

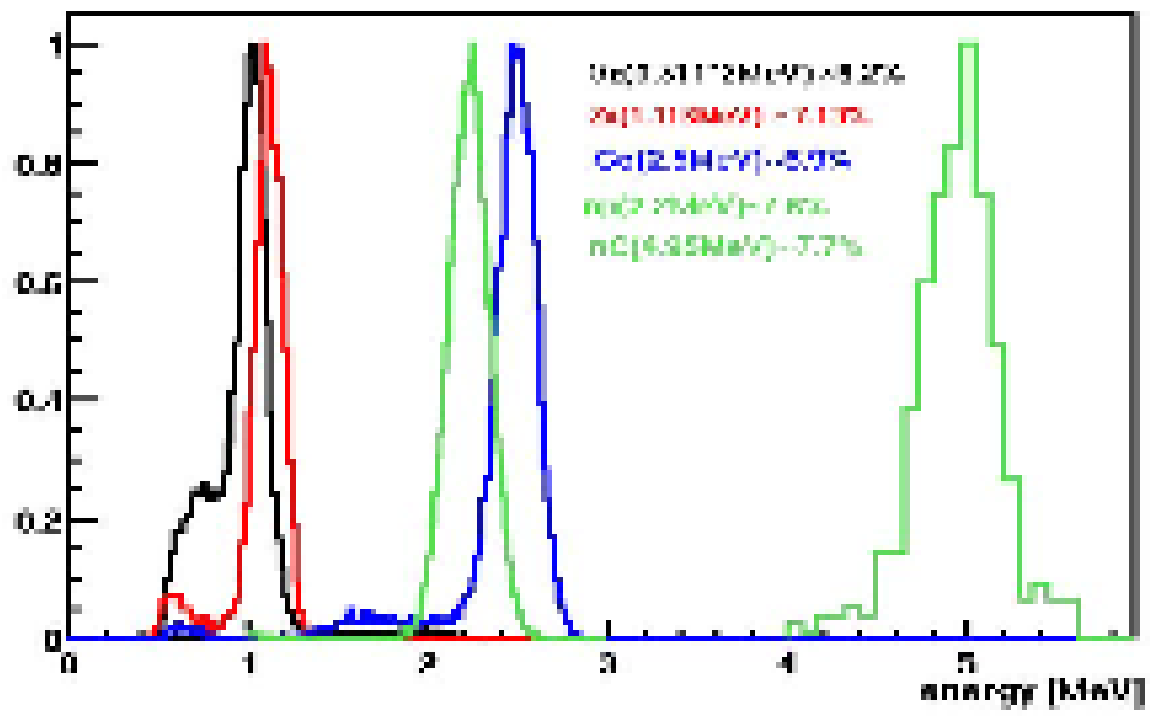

Fig.-8: Energy distribution for various radio active sources introduced into the balloon center and neutron absorption signal.

Fig.-9 shows reconstructed position distribution for $^{60}Co$ source deployed at z=0, $\pm$2, and $\pm$4m, where 0m is center of the balloon. The width includes the flight length of the two γ-rays. The width of the distribution is roughly $30\boldsymbol{cm}/\sqrt{\boldsymbol{E}(\boldsymbol{MeV})}$ per axis and the error of the center position can be tuned to be better than 5cm within the fiducial volume. Radio purity of the liquid scintillator was of major concern for our LS because the energy of reactor neutrino is low and the event rate is small and the signal could easily be swamped by backgrounds.

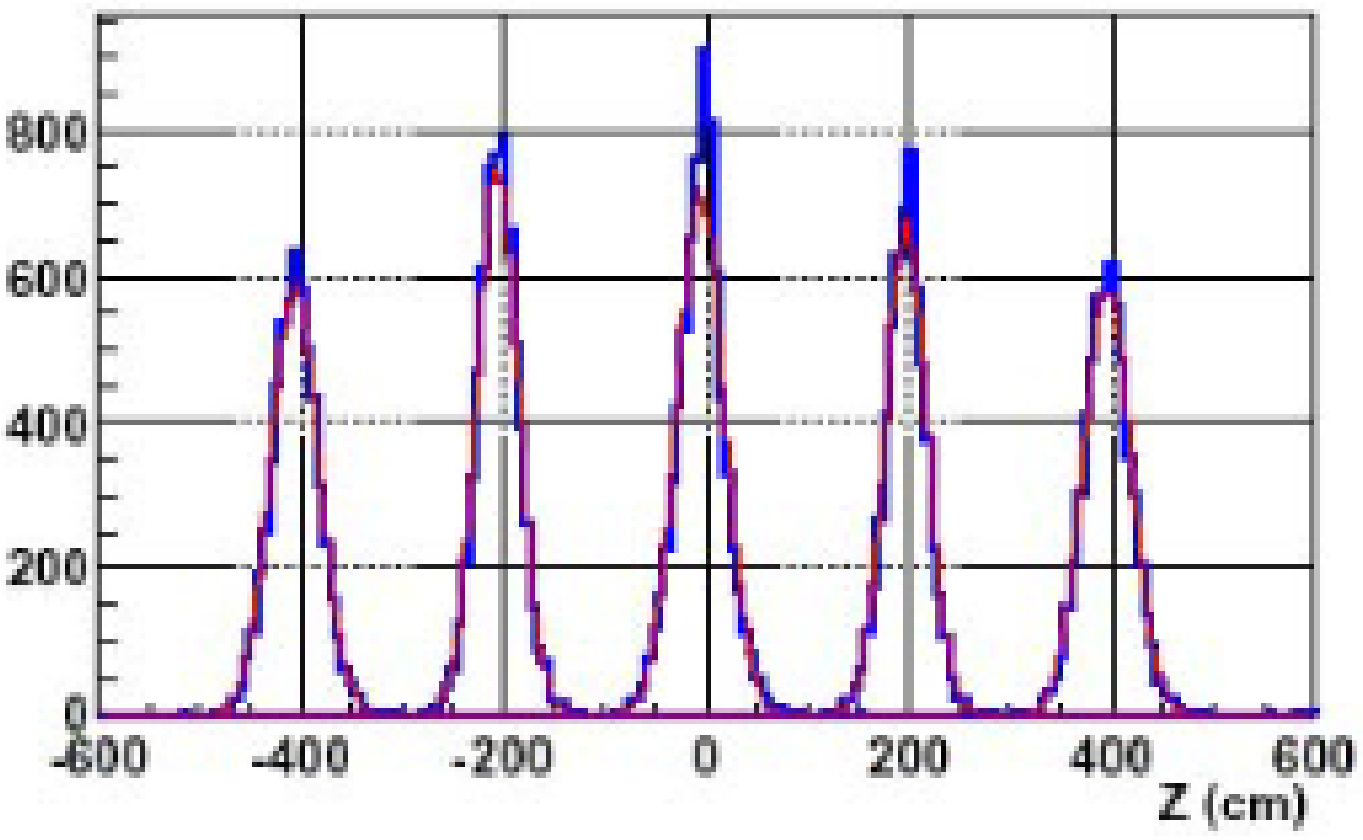

Fig.-9: Reconstructed position distribution for $^{60}Co$ source deployed at the position of 0, $\pm 2$, $\pm 4$ m, from the center. The width includes path length of the two ***g***-rays.

In order to make noise to signal ratio to be less than 1%, contamination level of $< 10^{-14} \boldsymbol{g/g}$ was required for U and Th and $< 10^{-12} \boldsymbol{g/g}$ for potassium. While the R&D phase, it turned out that the oils contain very small amount of U/Th/K, presumably because they are produced by a number of refining processes. To further purify, the water extraction was performed to both the LS and BO. The radio activity of the LS in the KamLAND detector was measured by using real signals. U contents was measured to be $3.5 \times 10^{-18} \boldsymbol{g/g}$ by tagging the successive decay of $^{214}Bi \rightarrow {}^{214}Po$ and Th contents was measured to be $5.2 \times 10^{-17} \boldsymbol{g/g}$ by tagging the successive decay of $^{212}Bi \rightarrow {}^{212}Po$, at the radius less than 5m, assuming radio equilibrium. The sensitivity of these measurements are the world record thanks to the huge amount of the 'specimen' and the measured concentrations are also the lowest value so far confirmed to have been achieved.

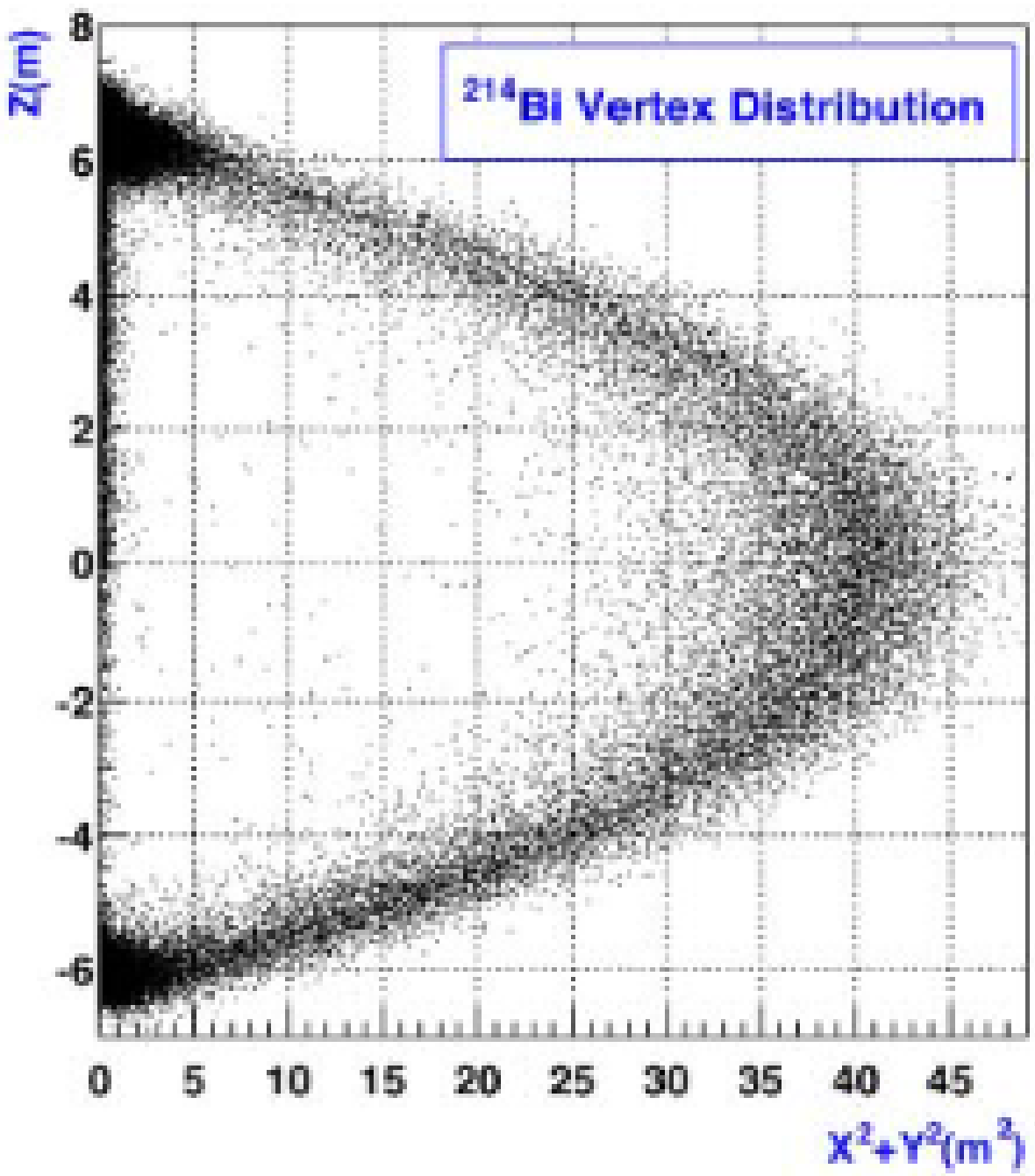

Fig.-10: Position dependence of the $^{214}Bi$ background. The horizontal axis is squared XY radius and the vertical axis is the z-coordinate.

Fig.-10 shows the position distribution of $^{214}$Bi-$^{214}$Po coincidence signals. The balloon shape is clearly seen because the balloon material contains much higher U contents (but still low if compared with regular materials) than LS. The high density region at the top and bottom part is due to the stainless steel structure of the balloon. The high density region at the vertical center is due to small thermometers installed, or $^{222}$Rn introduced when a calibration source is deployed every week. It can be seen that the background rate of the LS region is extremely low. For $^{40}$K, there is not useful coincidence signals to identify the source and only upper limit was measured from the energy distribution of the singles to be $<3.7\times10^{-16}\boldsymbol{g/g}$. Fig.-11 shows the energy distribution of the singles. For energies higher than 3 MeV, cosmic-ray spallation back grounds dominates. The $^{208}$Tl background comes from the γ-rays form PMT glasses or rock outside the detector and can be reduced further by setting harder fiducial cut if necessary. $^{210}$Pb is the decay product of the $^{222}$Rn which came in the LS while the LS was formulated. These values are low enough for reactor $\bar{\nu}_e$ detection.

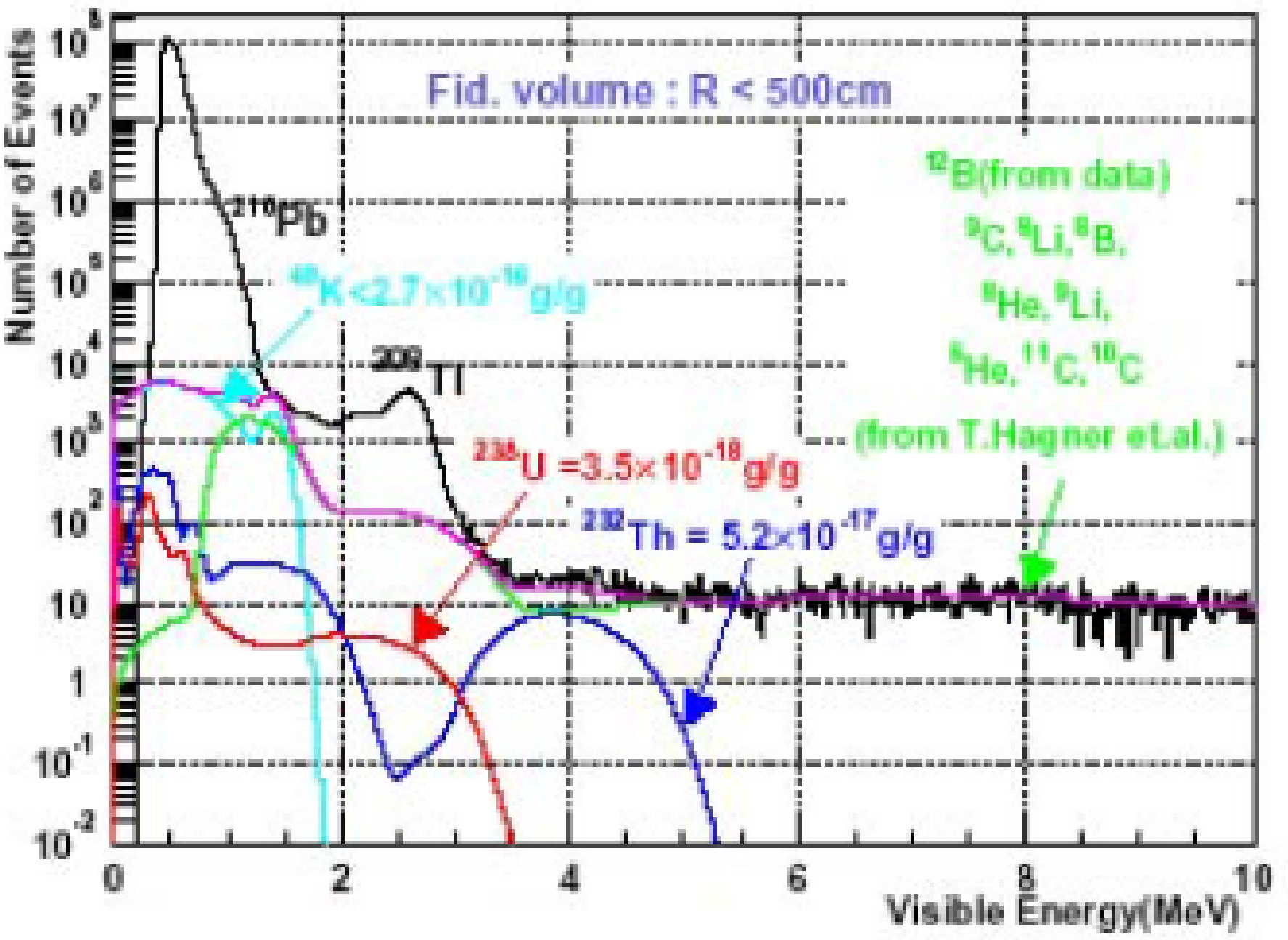

Fig.-11: Single event rate distribution. The fiducial volume is R<5m.
Spallation background is calculated from T.Hagner et al. [7]

## Summary

The KamLAND detector uses 1 kiloton of high quality LS as the active neutrino target. The LS is made up by normal-dodecane, pseudocumene and PPO. The total volume of the LS is 1,200m$^3$ which is the largest homogeneous liquid scintillator in the world. The preliminarily measured effective light output is 70% anthracene and the effective transparency is about 20m. The radio purity is extremely low, such as $3.5\times10^{-18}\boldsymbol{g/g}$ for uranium. Using this LS, reactor $\bar{\nu}_e$ s have been detected with high accuracy and very low background, and KamLAND found a large deficit of reactor neutrinos for the first time. KamLAND keeps data taking for several more years and the accuracy of the measurement is expected to significantly improve and other physics outputs are expected to come out in the future making the full use of this LS.